\documentclass{phc-proc4-auth}
\usepackage{amssymb}
\usepackage[dvips]{graphicx}
\newcommand{\ped}[1]{\ensuremath{_{\rm #1}}}

\begin{document}
\begin{frontmatter}

\title{Directional point-contact spectroscopy of MgB$_2$ single crystals
    in magnetic fields: two-band superconductivity and critical fields}

\author[To]{R.S. Gonnelli \corauthref{cor}}
\corauth[cor]{Corresponding author.}
\ead{renato.gonnelli@polito.it},
\author[To]{D. Daghero},
\author[To]{G.A. Ummarino},
\author[To]{Valeria Dellarocca},
\author[To]{A. Calzolari},
\author[Ru]{V.A. Stepanov},
\author[ETH]{J. Jun},
\author[ETH]{S.M. Kazakov},
\author[ETH]{J. Karpinski}

\address[To]{INFM - Dipartimento di Fisica, Politecnico di Torino,
Corso Duca degli Abruzzi 24, 10129 Torino, Italy}

\address[Ru]{P.N. Lebedev Physical Institute, Russian Academy of
Sciences, Leninsky Pr. 53, 119991 Moscow, Russia}

\address[ETH]{Solid State Physics Laboratory, ETH, CH-8093 Zurich,
Switzerland}

\begin{abstract}
The results of the first directional point-contact measurements in
MgB$_2$ single crystals, in the presence of magnetic fields up to
9 T either parallel or perpendicular to the $ab$ planes, are
presented. By applying suitable magnetic fields, we separated the
partial contributions of the $\sigma$ and $\pi$ bands to the total
Andreev-reflection conductance. Their fit with the BTK model
allowed a very accurate determination of the temperature
dependency of the gaps ($\Delta\ped{\sigma}$ and
$\Delta\ped{\pi}$), that resulted in close agreement with the
predictions of the two-band models for MgB$_2$. We also obtained,
for the first time with point-contact spectroscopy, the
temperature dependence of the (anisotropic) upper critical field
of the $\sigma$ band and of the (isotropic) upper critical field
of the $\pi$ band.
\end{abstract}

\begin{keyword}
point-contact spectroscopy; magnesium diboride; critical fields
\end{keyword}
\end{frontmatter}
Point-contact spectroscopy (PCS) has proved particularly useful in
investigating the two-band system MgB$_2$ \cite{twoband,Brinkman},
since it allows measuring both the $\sigma$- and $\pi$-band gaps
at the same time. In this paper we show how this technique,
combined with the use of magnetic fields and applied to MgB$_2$
single crystals, has allowed us to measure with the greatest
accuracy the temperature dependence of the two gaps
($\Delta\ped{\pi}$ and $\Delta\ped{\sigma}$), and of the upper
critical fields of the two bands.

\begin{figure}[t]
\begin{center}
\includegraphics[keepaspectratio, width=0.9\columnwidth]{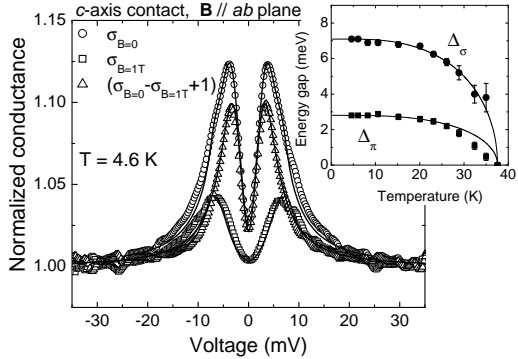}
\end{center}
\caption{Normalized conductance curve of a $c$-axis contact, with
no magnetic field (circles) and in a magnetic field of 1 T
parallel to the $ab$-plane (squares). Triangles: difference
between the previous two curves (vertically shifted by 1). Solid
lines are the best-fitting BTK curves. Inset: temperature
dependence of $\Delta\ped{\sigma}$ and $\Delta\ped{\pi}$ (symbols)
compared to BCS-like curves.} \label{fig:1}
\end{figure}

The high-quality, plate-like single crystals we used for our
measurements were produced at ETH (Zurich) by using the growth
technique described elsewhere\cite{Karpinski}. The point contacts,
that resulted always in the ballistic regime, were made by using
either a small spot of silver conductive paint or a small piece of
indium placed on the upper flat surface of the crystals or on
their side (so as to inject the current mainly parallel or
perpendicular to the $c$ axis, respectively). A typical
conductance curve of a $c$-axis contact, measured at $T=4.6$~K and
normalized to the normal-state conductance, is reported in
Fig.\ref{fig:1} (circles). It clearly shows the typical features
due to Andreev reflection at the normal metal-superconductor
interface, with conductance maxima at $V\approx \pm 3.5$~mV and a
smooth shoulder at $V\approx \pm 7.2$~mV, clearly related to
$\Delta\ped{\pi}$ and $\Delta\ped{\sigma}$, respectively. The
solid line superimposed to the experimental data is the
best-fitting curve obtained with the BTK model generalized to the
case of two bands, i.e. by expressing the total normalized
conductance as
$\sigma=w\ped{\pi}\sigma\ped{\pi}+(1-w\ped{\pi})\sigma\ped{\sigma}$,
where $\sigma\ped{\pi}$ and $\sigma\ped{\sigma}$ are the partial
normalized conductances of the two bands and $w\ped{\pi}$ is the
weight of the $\pi$ band, that depends on the angle $\phi$ between
the injected current and the $ab$ planes \cite{Brinkman}. Due to
the large number of free fitting parameters (the gaps
$\Delta\ped{\pi}$ and $\Delta\ped{\sigma}$, the barrier parameters
$Z\ped{\pi}$ and $Z\ped{\sigma}$, the broadening parameters
$\Gamma\ped{\pi}$ and $\Gamma\ped{\sigma}$ plus the weight
$w\ped{\pi}$), this fitting procedure gives a rather large
uncertainty on the gap values.
Reducing this uncertainty is however possible by applying a
suitable magnetic field (i.e. $B\simeq 1$~T at 4.6 K) that
completely destroys the superconductivity in the $\pi$ band
without affecting the $\sigma$ band \cite{nostroPRL}. The
resulting normalized conductance curve (squares in
Fig.\ref{fig:1}) can be simply represented as
$\sigma=w\ped{\pi}+(1-w\ped{\pi})\sigma\ped{\sigma}$ and can thus
be fitted with the standard BTK model, with only three fitting
parameters (plus $w\ped{\pi}$). The best-fit curve reported in
Fig.\ref{fig:1} was obtained with $w\ped{\pi}=0.98$,
$\Delta\ped{\sigma}$=7.09~meV (in very good agreement with the
prediction of the two-band model \cite{twoband,Brinkman}),
$Z\ped{\sigma}$=0.6, and $\Gamma\ped{\sigma}$=1.7~meV. To get the
values of the corresponding quantities relevant to the $\pi$ band,
we subtracted the curve measured in 1~T from that measured in zero
field. The resulting curve is reported, shifted by 1, in
Fig.\ref{fig:1} (triangles). It was fitted with a function of the
form $\sigma=w\ped{\pi}(\sigma\ped{\pi}-1)$ where $w\ped{\pi}$ is
kept equal to 0.98. The resulting best-fit parameters are:
$\Delta\ped{\pi}$=2.8~meV (very close to the theoretical value
\cite{twoband,Brinkman}), $Z\ped{\pi}$=0.6, and
$\Gamma\ped{\pi}$=2.0~meV. By using the same approach we were able
to determine the complete temperature dependence of the two gaps,
$\Delta\ped{\pi}$ and $\Delta\ped{\sigma}$, with unprecedented
accuracy \cite{nostroPRL}. The results are reported in the inset
of Fig.\ref{fig:1} (symbols) together with the BCS-like curves
with gap ratio $2\Delta/k\ped{B}T\ped{c}$ equal to 1.73 and 4.38,
respectively. The comparison shows that, at $T\!\!>$25~K,
$\Delta\ped{\pi}$ is lower than the BCS value, as predicted by the
two-band model \cite{twoband,Brinkman}.
\begin{figure}[t]
\vspace{-2mm}
\begin{center}
\includegraphics[keepaspectratio, width=\columnwidth]{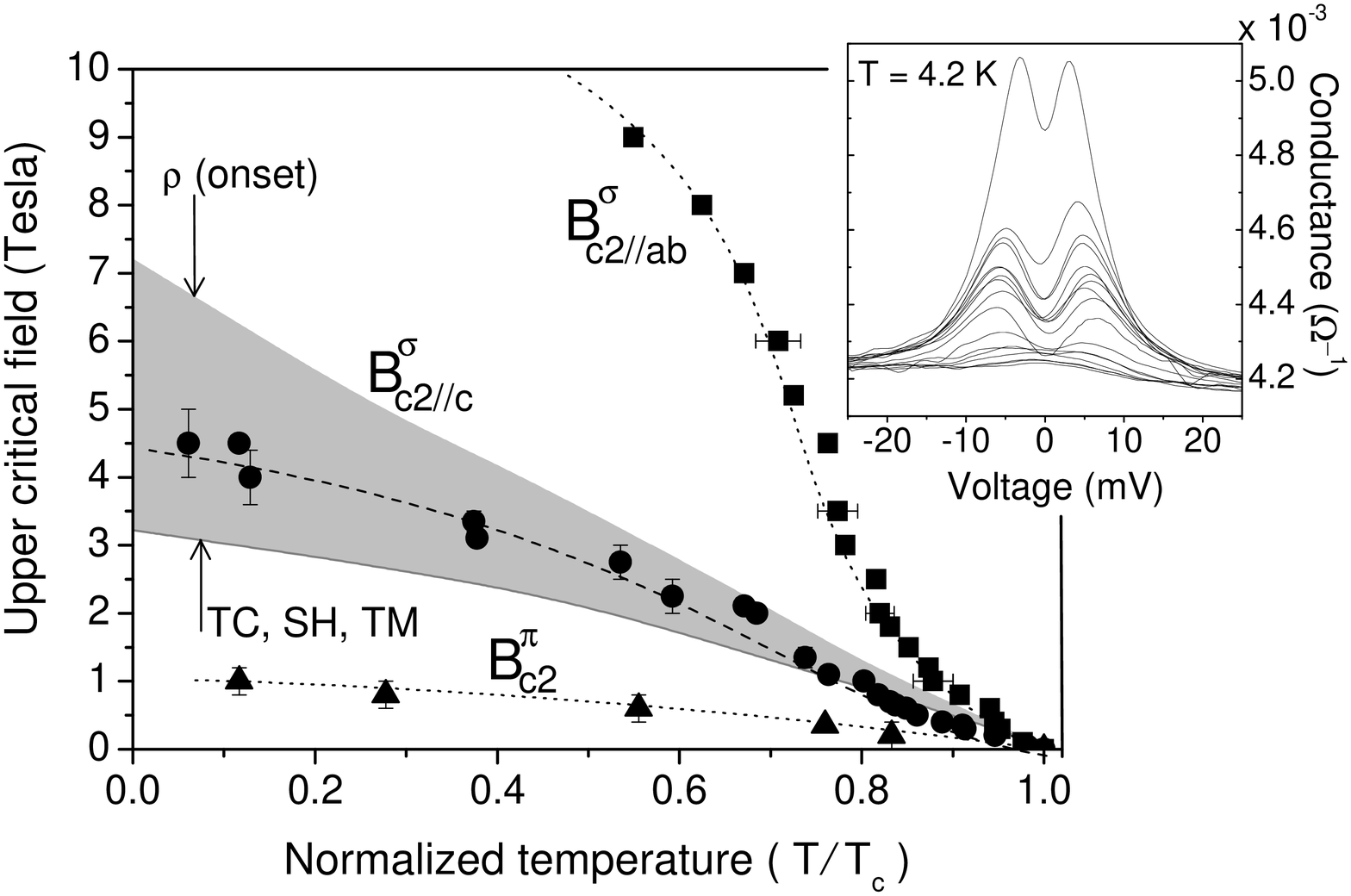}
\end{center}
\caption{Upper critical fields $B_{c2\parallel ab}^{\sigma}$
(squares), $B_{c2\parallel c}^{\sigma}$ (circles) and
$B_{c2}^{\pi}$ (triangles) vs. temperature. The shaded region is
upper-bounded by the values of $B_{c2\parallel c}^{\sigma}$ given
by the onset of the resistive transition \cite{Welp} and
lower-bounded by those given by thermal conductivity (TC)
\cite{Sologubenko}, torque magnetometry (TM) \cite{Angst}, and
specific heat (SH) measurements \cite{Welp}. Dotted lines are only
guides to the eye. Inset: conductance curves of a $ab$-plane
In/MgB$_2$ junction at 4.2~K in a magnetic field of increasing
intensity (up to 7 T) parallel to the $c$ axis. }\label{fig:2}
\end{figure}

An example of the effect of the magnetic field on the conductance
curves of a point contact (in particular, an $ab$-plane In/MgB$_2$
junction) is reported in the inset of Fig.\ref{fig:2}. The
magnetic field that destroys the superconductivity in the $\pi$
band, but leaves $\Delta\ped{\sigma}$ practically unchanged, is of
course the upper critical field of the $\pi$ band,
$B\ped{c2}^{\pi}$. In agreement with the isotropic character of
the $\pi$ bands, we found that $B\ped{c2}^{\pi}$ is independent of
the direction of the magnetic field \cite{nostroPhysicaC}. Its
temperature dependence is reported in the main panel of
Fig.~\ref{fig:2} (solid triangles). The same figure also reports
the values of the upper critical field of the $\sigma$ band, that
instead depends on whether $\mathbf{B}\!\!\parallel\!\! c$ or
$\mathbf{B}\!\!\parallel\!\! ab$. For both directions and at $T
\gtrsim 0.8 T\ped{c}$ the measured critical fields agree rather
well with those given by bulk-sensitive techniques such as torque
magnetometry (TM) \cite{Angst}, thermal conductivity (TC)
\cite{Sologubenko} and specific heat (SH) \cite{Welp}. At lower
$T$ our fields $B\ped{c2\parallel c}^{\sigma}$ and
$B\ped{c2\parallel ab}^{\sigma}$ are greater than those determined
by TM, TC and SH, but lower than the fields which mark the onset
of the resistive transition (see upper limit of the shaded region
in Fig.\ref{fig:2}), recently identified with the critical field
$B\ped{c3}$ \cite{Welp}. Even if PCS is a surface-sensitive
technique, this behavior cannot be simply due to surface
nucleation of superconductivity at a field $B\ped{0}$
($B\ped{c2}\!\!<\!\!B\ped{0}\!\!<\!\!B\ped{c3}$) because the
measured magnetoresistivity of In electrode is always much lower
than that of MgB$_2$ crystals \cite{Hurault}. The present results
might thus suggest the occurrence of some intrinsic effects, maybe
related to slight modifications of the superconducting properties
at the MgB$_2$ surface.

\end{document}